%% file: draft.tex
\documentclass[a4paper,11pt]{amsart}

\usepackage{globaldef}
\input{definitions.tex}

\bibliography{bib-database.bib}

\usepackage{amsaddr} 

\begin{document}


\title{\(Q\)-Boson model and relations with integrable hierarchies}

\author{Thiago Araujo}

\address{\noindent 
Instituto de Física Teórica, UNESP-Universidade Estadual Paulista, \\
Rua Dr. Bento T. Ferraz 271, Bl. II, São Paulo 01140-070, SP, Brazil
\\ \& \\
Instituto de Física, Universidade de São Paulo,\\
Rua do Matão, Travessa 1371, 05508-090 São Paulo, SP. Brazil
}

\email{\texttt{\href{tr.araujo@unesp.br}{tr.araujo@unesp.br}}}

\begin{abstract}
This work investigates the intricate relationship between the q-boson
model, a quantum integrable system, and classical integrable systems
such as the Toda and KP hierarchies. Initially, we analyze scalar
products of off-shell Bethe states and explore their connections to
tau functions of integrable hierarchies. Furthermore, we discuss
correlation functions within this formalism, examining their
representations in terms of tau functions, as well as their Schur
polynomial expansions.
\end{abstract}

\date{\today}
\keywords{Integrability, Bethe states, Schur, Hall-Littlewood, Toda, KP}
\subjclass[2020]{82B20, 82B23}

\maketitle

\setcounter{tocdepth}{1}
\tableofcontents

\section{Introduction}

The study of connections between quantum and classical
exactly solvable models is an important research program aimed at
elucidating the underlying structure of integrable systems. This
research program has yielded fruitful insights, as evidenced
by~\cite{Its:1992bj, Foda:2009zz, Alexandrov:2011aa, Araujo:2021ghu},
to name a few. The present work is situated within this research field.

Here, we examine the emergence of classical integrable structures
within correlation functions of the q-boson system. This quantum
integrable system describes q-deformed bosons confined to a
one-dimensional chain~\cite{Bogoliubov:1992, Bogoliubov:1997soj,
  Bogoliubov2005}.  Interestingly, this model is closely related to
the AL (Ablowitz-Ladik) model, which is an integrable discretization
of the nonlinear Schrödinger equation.

In~\cite{Bogoliubov2005}, the author shows that the \(q = 0\)
limit of the q-boson, known as the \emph{phase model}, is associated
with the enumeration of plane partitions. This connection allows for a
Schur polynomials expansion for the Bethe states. Subsequently, this
result was extended to \(q > 0\) in~\cite{Tsilevich:2006}, where the
Bethe states are expressed in terms of Hall-Littlewood polynomials.

The authors of~\cite{Foda:2008hn, Wheeler:2010vmq} utilize a rich
set of dualities between the ring of symmetric functions and
\(q\)-deformations of the Heisenberg algebra~\cite{Jing1991, Jing1995}
to explore the combinatorial properties of this spin chain. As a
consequence, it has been shown that the scalar products of off-shell
Bethe states are restricted tau functions of the KP
(Kadomtsev-Petviashvili) hierarchy.

This underlying combinatorial structure underscores the significance
of this model in various physical phenomena, notably in the context of
crystal melting and string theory~\cite{Okounkov:2003sp,
  Saulina:2004da}.  The application of the q-boson model in these
contexts has been demonstrated in~\cite{Sulkowski:2008mx}, with the
string data recoverable as the chain length approaches infinity.

This paper builds upon these developments and aims to extend some of
these results. Firstly, we investigate the connections between the
scalar product and tau functions, exploring them in a wider context,
particularly in the case of the Toda
hierarchy~\cite{Takasaki:2018wsv}. It is important to note that both
the AL and the KP hierarchies are reductions of the Toda
hierarchy. This paper takes steps towards resolving this problem,
which is instrumental in understanding the finer details of these
relationships.

Additionally, we explore other correlation functions within this model
and examine how they fit into the classical integrable context. It is
noteworthy that even when the object is not a tau function itself, it
may have an interesting expansion in terms of different types of
Schur polynomials, highlighting some of its combinatorial
properties\footnote{Results involving Schur and Hall-Littlewood
polynomials have been confirmed using the
software~\cite{Araujo:2024piv}.}. We also investigate some of these
expansions in this paper.

In Section 2, we provide a review of the phase model and
q-bosons. This section serves to establish notation and emphasize the
key aspects relevant to our analysis.  Section 3 delves into the
analysis of the phase model and its connections with the KP and Toda
integrable hierarchies. We explore the general expansion of a phase
model correlation function that yields a Toda system solution.  In
Section 4, we shift our focus to the q-bosons. We discuss a
determinantal formula for the scalar product of two off-shell Bethe
states and its expansion in terms of big, supersymmetric Schur
functions, as well as in terms of Kostka-Foulkes polynomials.
Finally, in Section 5, we conclude with a discussion of our results
and an overview of open problems.

\section{Phase Model and q-Bosons}

This section introduces the \emph{phase} and \emph{\(q\)-boson}
models~\cite{Bogoliubov:1992, Bogoliubov:1997soj, Bogoliubov2005,
  Tsilevich:2006}. It serves as a comprehensive review of existing
literature, offering insights into established concepts.  While the
content does not introduce novel ideas, its presentation adds value,
particularly from a pedagogical standpoint.  In structuring this
discussion, we adopt certain conventions from~\cite{Wheeler:2010vmq},
while aligning our presentation style more closely with that
of~\cite{Tsilevich:2006}.


\subsection{Phase model}
Consider the \((M+1)\) set of operators \(\{\phi_i,
\phi_i^\dagger,\mathcal{N}_i\}_{i=0}^M\) such that
\begin{equation}
 [\mathcal{N}_i, \phi_j] = - \phi_i \delta_{i,j} \quad
 [\mathcal{N}_i, \phi_j^\dagger] =  \phi_i^\dagger \delta_{i,j}  \quad 
 [\phi_i, \phi_j^\dagger] =  \pi_i \delta_{i,j}  
\end{equation}
where \(\pi_i =(\ket{0} \bra{0})_i\) is the vacuum projection operator.
These operators can be written as
\begin{equation}
\phi = \sum_{n\geq 0}\ket{n}\bra{n+1}\quad 
\phi^\dagger = \sum_{n\geq 0}\ket{n+1}\bra{n} \quad 
N = \sum_{n\geq 0}n\ket{n}\bra{n}\; ,
\end{equation}
and it is easy to see that \(\phi^\dagger \phi = \bm{1} - \ket{0}
\ket{0}\) and \(\phi\phi^\dagger = \bm{1}\).

The Hamiltonian is given by
\begin{equation}
  H = - \frac{1}{2} \sum_{n =0}^M \left(\phi_n^\dagger \phi_{n+1}
  + \phi_n \phi_{n+1}^\dagger \right) + \bar{\mathcal{N}}\; ,
\end{equation}
where \(\bar{\mathcal{N}} = \sum_{i=0}^M \mathcal{N}_i\) is the total
number operator, and we also impose periodic boundary conditions
\(\phi_{M+1} = \phi_0\) and \(\phi_{M+1}^\dagger = \phi_0^\dagger\).

These operators appear in the context of quantum optics, and for this
reason, this model is referred to as the \emph{phase model}. It corresponds to
the strongly correlated limit of the \(q\)-bosons
model~\cite{Bogoliubov:1997soj}, which we will define shortly.


\subsubsection{Representation}
The representation of the phase model algebra is constructed using
the vacuum state defined by \(\ket{0}_i\) by \(\phi_i\ket{0}_i =
0\). In this context, the state with \(n_i\) bosons (oscillators) is
given by \(\ket{n_i}_i = (\phi_i^\dagger)^{n_i} \ket{0}_i\).

Given the vacuum \(\ket{\bm{0}} = \ket{0}_0\otimes \ket{0}_1
\otimes \cdots \otimes  \ket{0}_M\), the Fock space is defined as 
\begin{equation}
  \mathcal{F} = \bigotimes_{i=0}^M \mathcal{F}_i = \mathrm{span}
  \left\{\ket{\vec{n}} = \ket{n_0}\otimes \ket{n_1} \otimes \cdots
  \otimes \ket{n_M} \ | \ n_i \in \mathbb{N} \right\}\; ,
\end{equation}
where the states \(\ket{\vec{n}}\) are defined as 
\begin{equation}
  \ket{\vec{n}} = \ket{n_0}\otimes \ket{n_1} \otimes \cdots \otimes \ket{n_M} 
 \equiv  (\phi_0^\dagger)^{n_0} (\phi_1^\dagger)^{n_1} \cdots  (\phi_M^\dagger)^{n_M} \ket{\bm{0}} \; ,
\end{equation}
with \(\phi_i^\dagger \equiv \bm{1} \otimes \cdots \otimes
\phi_i^\dagger \otimes \cdots \otimes \bm{1}\).  Moreover, it is easy
to see that these states are normalized, that is
\(\bracket{\vec{n}}{\vec{m}}=\delta_{\vec{n}, \vec{m}}\).  Finally,
the actions of the operators \(\mathcal{N}_i\) and \(\pi_i\) are
\begin{equation}
    \pi_i\ket{\vec{n}}  = \delta_{n_i, 0} \ket{\vec{n}} \qquad 
    \mathcal{N}_i\ket{\vec{n}} = n_i \ket{\vec{n}}\; .
\end{equation}
 
Given a state \(\ket{\vec{n}} = \ket{n_0, n_1, \dots, n_M}\), we
associate a partition \( \lambda = (1^{n_1} 2^{n_2} \cdots
M^{n_M})\).  It's worth noting that this correspondence is not unique,
as the partition \(\lambda\) does not account for the number of
particles \(n_0\).  If the total number of particles \(N\) is known,
we can determine \(n_0 = N - \ell(\lambda)\), where \(\ell(\lambda)\)
represents the number of rows in the Young diagram defined by
partition \(\lambda\).  This aspect is crucial because, in our
subsequent considerations, the \(N\) particle sector remains fixed
owing to the integrability of the model. Consequently, the value of
\(n_0\) becomes known once we specify the partition \(\lambda\).

Finally, based on the correspondence we mentioned in the introduction,
Wheeler~\cite{Wheeler:2010vmq} defines a map \(\mathcal{M}_\psi:
\mathcal{F} \to \mathcal{F}^{(0)}_\psi\), where
\(\mathcal{F}^{(0)}_\psi\) denotes the Fock space of charged free
fermions constructed from the neutral (fermionic) vacuum.


\subsubsection{Bethe Ansatz}
The \(L\)-matrix is given by 
\begin{equation}
  L_{an} = 
\begin{pmatrix}
x^{ - 1/2} & \phi_n^\dagger \\ \phi_n & x^{1/2}
\end{pmatrix}_a\; , 
\end{equation}
where \(x \in \mathbb{S}^1\). We also have the monodromy matrix
\begin{equation}
  T_a(x) = L_{aM}(x) \cdots L_{a0}(x) = 
\begin{pmatrix}
A(x) & B(x) \\ C(x) & D(x)
\end{pmatrix}_a\; .
\end{equation}

With these expressions, one can finally build the Bethe states
\begin{equation}
  \ket{\Psi(y_1, \dots, y_N)} = \prod_{j=1}^N \mathbb{B}(y_j) \ket{\bm{0}} \qquad 
  \bra{\Psi(y_1, \dots, y_N)} = \bra{\bm{0}}\prod_{j=1}^N \mathbb{C}(y_j) 
\end{equation}
where \(\mathbb{B}(y) = y^{M/2} B(y)\) and \(\mathbb{C}(y) = y^{M/2} C(1/y)\).

When the coordinates \(\{ y_j \ | \ j =1, \dots , N\}\) satisfy the Bethe equations
\begin{equation}
\label{eq:bethe_eq}
  y^{N + M}_i = (-1)^{N-1} \prod_{\substack{j = 1 \\ j \neq i}}^N y_j\; , \qquad i = 1, \dots, N\; , 
\end{equation}
we say that the Bethe states are \emph{on-shell}; otherwise, we have
\emph{off-shell} states. In what follows, we will only consider
\emph{off-shell} states.


\subsection{q-Bosons}
The set of operators \(\{b_i, b_i^\dagger,\mathcal{N}_i\}_{i=0}^M\) constitute
\(M+1\) independent q-boson algebras defined by
\begin{equation}
[\mathcal{N}_i, b_j^\dagger]=\delta_{i,j} b_i^\dagger\; , \quad 
[\mathcal{N}_i, b_j]=-\delta_{i,j}b_i\; , \quad
[b_i, b_j^\dagger]= \delta_{i,j} q^{-2\mathcal{N}_i}  \equiv \delta_{i,j} Q^{\mathcal{N}_i}\; , 
\end{equation}
where we denote the deformation parameter as \(Q = q^{-2}\).

The q-boson model is characterized by its Hamiltonian
\begin{equation}
  \mathcal{H} = -\frac{1}{2} \sum_{n=0}^M
  \left(b_n^\dagger b_{n+1} + b_n b_{n+1}^\dagger \right) + \bar{\mathcal{N}}\; ,
\end{equation}
where \(\bar{\mathcal{N}} = \sum_{i=0}^M \mathcal{N}_i\), and we also
impose periodic boundary conditions \(b_{M+1} = b_0\) and
\(b_{M+1}^\dagger = b_0^\dagger\).

In the limit \(Q \to 1\), the q-bosons behave as ordinary bosons,
while the limit \(Q \to 0\) (\(q \to \infty\)) corresponds to the
phase model discussed earlier.


\subsubsection{Representation}
Let us define the \(i\)-th vacuum \(\ket{0}_i\) such that
\(b_i\ket{0}_i = 0\). The representation of this Hilbert space,
denoted by \(\mathcal{F}_i^Q\), is defined by the states
\(\ket{n_i}_i \propto (b_i^\dagger)^{n_i} \ket{0}_i\).

Given the vacuum \(\ket{\bm{0}} = \ket{0}_0\otimes \ket{0}_1
\otimes \cdots \otimes  \ket{0}_M\),
the Fock space is defined as 
\begin{equation}
  \mathcal{F}_{Q} = \bigotimes_{i=0}^M \mathcal{F}_i^Q = 
  \mathrm{span}\left\{\ket{\vec{n}} = \ket{n_0}\otimes \ket{n_1} \otimes \cdots
  \otimes \ket{n_M} \ | \ n_i \in \mathbb{N} \right\}\; ,
\end{equation}
where the actions of the operators \(\{b_i, b_i^\dagger\}\) are given
by the following relations
\begin{subequations}
\begin{alignat}{3}
    & b_0\ket{n_0} =(1 - \delta_{0, n_0}) \frac{(1 - Q^{n_0})}{(1 - Q)^{1/2}}\ket{n_0 - 1}
    &\hspace{0.5cm}& b_0^\dagger \ket{n_0} =  \frac{1}{(1 - Q)^{1/2}}\ket{n_0 + 1}  \\
    & b_i\ket{n_i} = \frac{(1 - \delta_{0, n_i})}{(1 - Q)^{1/2}}\ket{n_i - 1}
    && b_i^\dagger \ket{n_0} =  \frac{(1 - Q^{n_i+1})}{(1 - Q)^{1/2}}\ket{n_0 + 1}\quad i\neq 0\; .
\end{alignat}
\end{subequations}

Consequently, the states \(\ket{\vec{n}} \in \mathcal{F}_Q\) are
\begin{equation}
  \ket{\vec{n}} = \ket{n_0}\otimes \ket{n_1} \otimes \cdots \otimes \ket{n_M} 
 : =  (b_0^\dagger)^{n_0} (b_1^\dagger)^{n_1} \cdots  (b_M^\dagger)^{n_M} \ket{\bm{0}} \; ,
\end{equation}
where we write \(b_i^\dagger \equiv \bm{1} \otimes  \cdots \otimes
b_i^\dagger \otimes \cdots \otimes \bm{1}\).
Moreover, this space has an inner product that satisfies
\begin{equation}
  \bracket{\vec{n}}{\vec{m}} = \frac{[m_0]!}{\prod_{i=1}^M[m_i]!} \delta_{\vec{n}\vec{m}}\qquad 
    [n]! =
    \left\{
    \begin{array}{ll}
    \prod_{i=1}^n(1 - Q^i) & \textrm{if}\ \ n\neq 0 \\
    1 & \textrm{otherwise}
  \end{array}\right.\; .
\end{equation}

Similar to the phase model, we can associate to a given states
\(\ket{\vec{n}} = \ket{n_0, n_1, \dots, n_M}\), a Young diagram
\(\lambda = (1^{n_1} 2^{n_2} \cdots M^{n_M})\). It is useful to define
a proportionality factor relating these two objects
\begin{equation}
  \ket{\lambda}_Q  = b_\lambda(Q) \ket{\vec{n}}  \ \; , \qquad b_\lambda(Q) = \prod_i [p_i(\lambda)]! 
\end{equation}
where \(p_i(\lambda)\) denotes the number of parts of size \(i\) in
the partition. These partition states satisfy
\(\bracket{\lambda}{\mu}_Q = b_\lambda(Q) \delta_{\lambda,\mu}\). Once
again, we see that the correspondence is not unique, as the number of
oscillators at site \(i=0\) is completely ignored in the partition
notation. Finally, if the number of particles is fixed, then \(n_0 = N
- \ell(\lambda)\).


\subsubsection{Bethe Ansatz}
The \(L\)-operator for the q-boson is given by
\begin{equation}
  L_{an}(x, Q) =
  \begin{pmatrix}
    x^{-1/2} & (1 - Q)^{\frac{1}{2}} b_n^\dagger \\ (1 - Q)^{\frac{1}{2}} b_n & x^{1/2}
  \end{pmatrix}_a\; ,
\end{equation}
and the monodromy matrix is 
\begin{equation}
  T_a(x,Q) = L_{am}(x, Q)  \dots  L_{a0}(x, Q) = 
  \begin{pmatrix}
    A(x, Q) & B(x, Q) \\ C(x, Q) & D(x, Q)
  \end{pmatrix}_a\; .
\end{equation}

As before, the eigenstates of the Hamiltonian have the form
\begin{equation}
  \ket{\Psi(y_1, \dots, y_N; Q)} = \prod_{j=1}^N \mathbb{B}(y_j, Q) \ket{\bm{0}}\qquad 
  \bra{\Psi(y_1, \dots, y_N; Q)} = \bra{\bm{0}}\prod_{j=1}^N \mathbb{C}(y_j, Q) \; ,
\end{equation}
where \(\mathbb{B}(y, Q) = y^{M/2} B(y, Q)\) and \(\mathbb{C}(y, Q) =
y^{M/2} C(1/y, Q)\). When the parameters \(\{ y_j \ | \ j =1, \dots , N\}\)
satisfy the Bethe equations given by
\begin{equation}
  y^{N + M}_i =\prod_{\substack{j = 1 \\ j \neq i}}^N\frac{Q y_i - y_j}{y_i - Q y_j}\; , \qquad i = 1, \dots, N\; , 
\end{equation}
we have \emph{on-shell} Bethe states; otherwise, they are \emph{off-shell} states.

\section{Tau functions in the phase model}

This section explores the presence of integrable hierarchies in the
phase model. We demonstrate its relations with the Toda hierarchy tau
function and discuss some implications, particularly its connection to
a matrix model. We also argue that these results imply that the scalar
products in the model also serve as KP hierarchy tau functions,
consistent with the findings of
Wheeler~\cite{Wheeler:2010vmq}. Additionally, we highlight the
existence of correlation functions in this model that satisfy the KP
hierarchy equations.

Bogoliubov~\cite{Bogoliubov2005} has demonstrated that the scalar
product of two vectors in the \(N\)-particle sector of a chain with
length \(M+1\) is
\begin{equation}
\begin{split}
\label{eq:scalar}
  \mathcal{I}(N, M|\bm{x}, \bm{y}) & =
  \bra{\bm{0}} \prod_{i=1}^N \mathbb{C}(x_i) \prod_{j=1}^N \mathbb{B}(y_j) \ket{\bm{0}} \\ 
 & = \frac{ \det H(\bm{x},\bm{y})}{ \prod_{i<j}(x_i - x_j)(y_i - y_j)} \; ,
\end{split}
\end{equation}
where \(H\equiv [H_{ij}]_{i,j=1}^N\) is an \(N\times N\) matrix with components
\begin{equation}
\label{eq:h-matrix}
  H_{ij} = H(x_i, y_j) 
  =\frac{1 - (x_i y_j)^{ M + N}}{1 - x_i y_j }\; .
\end{equation}


\subsection{Toda tau functions}
We now argue that the scalar product defined above is a tau function
of the Toda hierarchy. Let us first write the function \(H(z,w)\) as
the geometric sum
\begin{equation}
\label{eq:h-exp}
  H(z,w) = \frac{1 - (zw)^{M+N}}{1 - zw} = \sum_{k=1}^{M+N} (zw)^{k-1} \; .
\end{equation}
Hence, we express the determinant of the \(H\) matrix as
\begin{equation}
  \det_{i,j} \left(H(x_i, y_j)\right) = \det_{i,j} \left( \sum_{k=1}^{M+N} x_i^{k-1} y_j^{k-1}\right)
\end{equation}
Furthermore, we can interpret this expression as the result of
multiplying an \(N\times (N+M)\) matrix \(\mathcal{X}\) by another
\((M + N)\times N\) matrix \(\mathcal{Y}\), which are given by
\begin{equation}
  \mathcal{X} = 
  \begin{pmatrix}
  x_1^0 & x_1^1 & \dots & x_1^{M+N-1} \\  
  x_2^0 & x_2^1 & \dots & x_2^{M+N-1} \\  
  \vdots \\
  x_N^0 & x_N^1 & \dots & x_N^{M+N-1} 
  \end{pmatrix}\quad \textrm{and} \quad 
  \mathcal{Y} = 
  \begin{pmatrix}
  y_1^0 & y_2^0 & \dots & y_N^0 \\  
  y_1^1 & y_2^1 & \dots & y_N^1 \\  
  \vdots & \vdots & & \vdots \\
  y_1^{N+M-1} & y_2^{M+N-1} & \dots & y_N^{M+N-1}
  \end{pmatrix}\; ,
\end{equation}
therefore 
\begin{equation}
  \det_{i,j} \left(H(x_i, y_j)\right)
  \equiv  \det_{i,j} \left( \mathcal{X}\mathcal{Y}\right)
  = \sum_{0 \leq \ell_{N} \leq \dots \leq \ell_1\leq N+M } \det_{ik}(x_i^{\ell_k}) \det_{ik}(y_j^{\ell_k})\; .
\end{equation}
If we now define \(\ell_j = \lambda_k - k + N\), and
using~(\ref{eq:scalar}), we have
\begin{equation}
\label{eq:scalar_exp}
\mathcal{I}(N, M|\bm{x}, \bm{y}) = \sum_{\lambda\subseteq [N,M]}
\frac{\det_{ik}(x_i^{\lambda_k - k + N})}{\Delta(\bm{x})} \frac{ \det_{ik}(y_j^{\lambda_k - k + N})}{\Delta(\bm{y})}
= \sum_{\lambda\subseteq [N,M]} s_\lambda(\bm{x}) s_\lambda(\bm{y}) \; ,
\end{equation}
where \(\Delta(\bm{x})\) and \(\Delta(\bm{x})\) are Vandermonde
determinants. This formula agrees with the Schur expansion defined
in~\cite{Bogoliubov2005}.

Define two sets of Miwa coordinates \(\bm{t} = (t_1, t_2, \dots)\) and
\(\bm{t}' = (t'_{-1}, t'_{-2}, \dots)\) as
\begin{equation}
  t_q = \frac{1}{q}\sum_{j=1}^N x_j^p\qquad 
  t'_{-q} = \frac{1}{q}\sum_{j=1}^N y_j^p\; , 
\end{equation}
where \(p_q(\bm{x}) = q t_q\) are power sums. One can write the inner
product in terms of these coordinates, that is
\begin{equation}
  \mathcal{I}(N, M|\bm{t}, \bm{t}') = \sum_{\lambda\subseteq [N,M]} s_\lambda(\bm{t}) s_\lambda(\bm{t}') \; .
\end{equation}
This expression is known to be a tau function for \(M, N\to \infty\).
As such, utilizing the free fermions representation~\cite{Alexandrov:2012tr}, it can be written as
\begin{equation}
  \lim_{M,N\to \infty}\mathcal{I}(N, M|\bm{t}, \bm{t}')
 = \bra{\bm{0}} e^{\bm{J}_+(\bm{t})} e^{-\bm{J}_-(\bm{t}')} \ket{\bm{0}}\; .
\end{equation}
It is a tau function of the Toda hierarchy with trivial element
\(\mathbb{1} \in GL(\infty)\), and it is nothing but the Cauchy's
identity
\begin{equation}
  \sum_{\lambda } s_{\lambda}(\bm{t}) s_{\lambda}(\bm{t}')
    = \exp \left( \sum_{m\geq1} m t_m t'_{-m} \right) \; .
\end{equation}

Bringing all these facts together, the truncation for finite \(M\) and
\(N\) also yields tau functions of the Toda hierarchy. More
specifically, according to~\cite{Alexandrov:2012tr, Kharchev:1991gd,
  Zabrodin:2010ii}, the truncation of the tau function corresponds to
the inclusion of a projection operator in the expectation value of the
tau function written in the fermionic representation.

As a final remark, we can also write
\begin{equation}
  H(z,w) = \sum_{k=0}^{M+N-1} h_k(zw)^{k} \; , 
\end{equation}
with \(h_k=1\) if \(k\in [0, M+N-1]\) and \(0\) otherwise. In this
case, one can define a diagonal \((N+M)\times (N+M)\) matrix
\(\mathcal{H} = \textrm{diag}(h_0, \dots, h_{M+N-1})\). Consequently,
if we repeat the arguments above, we find
\begin{equation}
\mathcal{I}(N, M|\bm{t}, \bm{t}') = \sum_{\lambda \subseteq [N,M]} h_\lambda s_\lambda(\bm{t}) s_\lambda(\bm{t}') \; .
\end{equation}
Here, \(h_\lambda\) is equal to \(1\) if \(\lambda \in [N,M]\), and it
is zero otherwise.

In this case, we find that this tau function is a trivial example of
the tau functions considered in~\cite{orlov:2001}. We anticipate that
in the analysis of more general correlation functions, the diagonal
terms \(h_\lambda\) will be more interesting. We will revisit this
discussion soon.


\subsubsection{Matrix Models}
It is also interesting to note the particular case when \(M\to
\infty\), but with a finite number of particles \(N\). In this case,
the partitions \(\lambda \in [N, \infty]\) satisfy the condition
\(\ell(\lambda) \leq N\). Therefore, the scalar
product~(\ref{eq:scalar_exp}) becomes
\begin{equation}
  \mathcal{I}_N(\bm{t}, \bm{t}')\equiv 
  \lim_{M\to \infty}\mathcal{I}(N, M|\bm{t}, \bm{t}')
  = \sum_{\substack{\lambda \\ \ell(\lambda) \leq N}} s_\lambda(\bm{t})  s_\lambda(\bm{t}') \; .
\end{equation}
From~\cite{Zabrodin:2010ii}, we know that this expression can be
written as the following integral
\begin{equation}
  \mathcal{I}_N(\bm{t}, \bm{t}') =
  \frac{1}{N!} \prod_{\ell=1}^N \oint_{\Gamma_\ell} \frac{dz_\ell}{2 \pi i z_\ell}
  e^{\xi(\bm{t}, z_\ell) - \xi(\bm{t}', z_{\ell}^{-1})} \Delta(z)\Delta(z^{-1})\; ,
\end{equation}
where
\begin{equation}
\xi(\bm{t}, z) = \sum_{k\geq 0} t_k z^k \qquad
\xi(\bm{t}', 1/z) = \sum_{k\geq 0} t'_{-k} z^{-k} \; . 
\end{equation}
See also~\cite{Kharchev:1991gd} for a detailed proof of this relation,
and~\cite{Orlov:2005} for other details. 

From the results of~\cite{Zabrodin:2010ii}, see citations therein, we
have an interesting consequence of this representation. Impose the
Bethe equations~(\ref{eq:bethe_eq}) to one set of variables, say \(x_j
= e^{-ip_j} \in \mathbb{S}^1\). Additionally, let us set \(\bm{t}' = -
\bm{t}^\star\). In this particular case, we have
\begin{equation}
\xi(\bm{t}, z) - \xi(\bm{t}', 1/z)  = 2 \ \textrm{Re}\left(\sum_k t_k z^k\right)\; .
\end{equation}
Then, the phase model is equivalent to an ensemble of \(N\) 2D Coulomb
particles on a circle. In this case, we find that the quantities
\(z_{\ell}\) are eigenvalues of a matrix \(U\).

Furthermore, according to Zabrodin~\cite{Zabrodin:2010ii}, see
citations therein, we also know that under the rescaling \(t_k \to
T_k/ \hbar\), \(t'_k \to T_{-k}/ \hbar\) and \(N = T_0/ \hbar\), we
obtain the dispersion limit tau function
\begin{equation}
  F_0(\bm{T}) = \log \mathcal{I}_{T_0}(\bm{T}, \bm{T}') + \mathcal{O}(\hbar)\; ,
\end{equation}
that is a free energy, from the viewpoint of the matrix integral
partition function.

It remains unclear how one can use this fact to determine properties
of the integrable model, but it might be possible to study the
analytic structure of the free energy \(F_0\) to gain some
understanding of the Bethe roots \(\bm{x}\). This problem is currently
under further investigation, and we hope to report new results
elsewhere.


\subsubsection{KP tau function}
Lastly, one may also observe that if we fix one set of coordinates,
say \(\bm{y}\), then we can write
\begin{equation}
 \lim_{M,N\to \infty}\mathcal{I}(N,M|\bm{t}) 
 = \sum_{\lambda} s_\lambda(\bm{y})  s_\lambda(\bm{t}) 
 \equiv \sum_{\lambda} c_\lambda(\bm{y})  s_\lambda(\bm{t}) \; ,
\end{equation}
with coefficients \(c_\lambda(\bm{y}) = \det (h_{\lambda_i-i
  +j}(\bm{y}))\). But now, it is trivial to notice that these are
Plücker coordinates in the Jacobi-Trudi form, as seen
in~\cite{Miwa2000, Alexandrov:2012tr}.

Hence, the following expression
\begin{equation}
 \mathcal{I}(N,M|\bm{t}) = \sum_{\lambda \subseteq [N,M]} s_\lambda(\bm{y})  s_\lambda(\bm{t}) \; ,
\end{equation}
is also a KP tau function, a fact that we already know
from~\cite{Wheeler:2010vmq}, where the author proved this statement
using the free fermions formalism.


\subsection{Correlation functions}
Bogoliubov has also shown, in~\cite{Bogoliubov2005}, that the
correlation functions
\begin{subequations}
\begin{equation}
\begin{split}
\label{eq:correlation}
  A_m(N, M|\bm{x}, \bm{y}\setminus \{y_N\})
  & = \bra{0} \prod_{j=1}^N \mathbb{C}(x_j)
  \prod_{k=1}^{N-1} \mathbb{B}(y_k) \phi_m^\dagger \ket{0}\\
  & =  \prod_{j=1}^N x_{j}^{M/2} \prod_{k=1}^{N-1} y_{j}^{M/2}
  \bra{0} \prod_{j=1}^N C(1/x_j) \prod_{k=1}^{N-1} B(y_k) \phi_m^\dagger \ket{0}
\end{split}
\end{equation}
can be written as
\begin{equation}
  A_m(N, M|\bm{x}, \bm{y}\setminus \{y_N\}) = 
  \frac{(-1)^{N-1}}{y_N^{(N-1)/2}} \prod_{j=1}^N x_{j}^{M/2}
  \prod_{k=1}^{N-1} y_{j}^{M/2}
  \left( \prod_{t<N} \frac{y_N - y_t}{y_t} \right)
  \frac{\det Q}{\det H} \mathcal{I}(N,M|\bm{x}, \bm{y})
\end{equation}
\end{subequations}
where \(Q\) is an \(N\times N\) matrix with components 
\begin{equation}
 Q_{jN} = x_j^{(M + N - 1- 2m)/2} \quad  \textrm{and} \quad 
 Q_{jk} = H_{jk} \; , 
\end{equation}
and \(H_{jk} = H(x_j, y_k)\) are the components of the matrix \(H\)
in~(\ref{eq:h-matrix}).  The components \(Q_{jN}\) are independent of
the coordinates \(y\); therefore, we cannot express the above
expression as a Toda hierarchy tau function.

We already know from~(\ref{eq:scalar}) that 
\begin{equation}
    \frac{\mathcal{I}(N,M|\bm{x}, \bm{y})}{\det H}=  \frac{1}{\Delta(x)\Delta(y)} \; ,
\end{equation}
then
\begin{equation}
  A_m(N, M|\bm{x}, \bm{y}\setminus \{y_N\})  \propto
  \frac{\det Q}{\Delta(x)} \equiv \mathcal{A}_m(N,M|\bm{x}, \bm{y})\; ,
\end{equation}
and we treat the coordinates \(\{ \bm{y} \}\) as a set of \(N\) fixed
parameters.

Furthermore, we define vector field \(\bm{F}(z) = (F_1, \dots, F_N)\),
where its components are given by
\begin{equation}
    F_j (z) = H(z, y_j) \quad \textrm{if} \ j \neq N\; , \qquad \textrm{and}\qquad 
    F_N (z)  = z^{(M + N - 1 - 2m)/2} \; .
\end{equation}
As before, we expand \(F_j(z)\), \(j\neq N\), as the geometric sum
\begin{equation}
  F_j(z) = \frac{1 - (y_j z)^{M + N}}{1 - y_j z} = \sum_{n=0}^{N + M - 1} (y_j z)^n
  \equiv \sum_{n=0}^{M + N -1} f_{j, n} z^n \; , \quad  f_{j, n} = y_j^n\; .
\end{equation}
We can also express the function \(F_N(z)\) in this form by setting
\(f_{N,n} = \delta_{n,(M - N - 1 - 2m)/2}\) and requiring \((M - N -
1)\) to be an even integer.

With these definitions, we conclude that
\begin{equation}
\mathcal{A}_m(N,M|\bm{x}, \bm{y}) =\frac{\det_{jk} F_j(x_k)}{\Delta(x)}\; .
\end{equation}
From the expansion 
\begin{equation}
  \det_{jk} F_j(x_k) = \det_{jk} \left(  \sum_{n=0}^{M + N -1} f_{j, n} x_k^n \right) \; ,
\end{equation}
and utilizing the Cauchy-Binet formula, we get
\begin{equation}
    \mathcal{A}_m(N,M|\bm{x}, \bm{y})
  = \sum_{0\leq \ell_N\leq \dots \leq \ell_1 \leq N+M}
  \frac{\det_{jk}(f_{j, \ell_k}) \det_{jk}(x_k^{\ell_j})}{\Delta(x)}\; . 
\end{equation}

We now use the definition of the Schur polynomials and the
Jacobi-Trudi expression of Plücker coordinates, as detailed
in~\cite{Alexandrov:2012tr}, leading to the following expression:
\begin{equation}
\mathcal{A}_m(N,M|\bm{x}, \bm{y}) =
\sum_{\lambda} c_\lambda(\bm{y}) s_\lambda(\bm{x}) \; ,
\end{equation}
where \(c_\lambda \equiv \det_{jk}(y_j^{\ell_k})\), \(\ell_k =
\lambda_k - k +N\).  Putting all these facts together, we conclude
that this expression is also a KP tau function.

This expression underscores the non-trivial nature of these tau
functions within the model. However, it also implies the existence of
other intriguing examples awaiting exploration. Let us now briefly
investigate other cases.


\subsubsection{Skew Schur polynomials expansion}
It has also been demonstrated in~\cite{Bogoliubov2005, Tsilevich:2006}
that the Bethe states exhibit a coordinate expansion
\begin{equation}
  \prod_{j=1}^N \mathbb{B}(x_j)\ket{\lambda}  =
  \sum_{\substack{\mu \supset \lambda \\ \mu \subseteq [N,M]}} s_{\mu/\lambda}(\bm{x})\ket{\mu}\qquad 
  \bra{\lambda} \prod_{j=1}^N \mathbb{C}(x_j) = \sum_{\substack{\mu \supset \lambda \\ \mu \subseteq [N,M]}} 
  s_{\mu/\lambda}(\bm{x})\bra{\mu}\; ,
\end{equation}
where \(s_{\mu/\lambda}\) are skew Schur polynomials~\cite{Macdonald:1998}.

Therefore, we can write the correlation function~(\ref{eq:correlation}) as 
\begin{equation}
\begin{split}
  A_m(N, M|\bm{x}, \bm{y}\setminus \{y_N\})
  & = \sum_\lambda \sum_{\substack{\mu \supset (m)\\ \mu \subseteq [N-1, M]}}
  s_{\lambda}(\bm{x}) s_{\mu/(m)}(\bm{y}\setminus\{y_N\})\\
  & = \sum_{\substack{\mu \supset (m)\\ \mu \subseteq [N-1, M]}}
  s_{\mu/(m)}(\bm{y}\setminus\{y_N\}) s_{\mu}(\bm{x})\; ,
\end{split}
\end{equation}
where we have used that \(\phi^\dagger \ket{\bm{0}} = \ket{(m)}\), and in the second line
we sum over all partitions, since \(s_{\mu/(m)} = 0\), \(\forall \) \(\mu \not \supset (m)\).
This expression also shows that the expression \(A_m\) can also be written as a tau function. 

More generally, let us consider the correlation functions 
\begin{equation}
\begin{split}
  A_{\lambda_1 \lambda_2}(N', N, M|\bm{x}, \bm{y}) & =
  \bra{\lambda_1} \prod_{j=1}^N \mathbb{C}(x_j)
  \prod_{k=1}^{N'} \mathbb{B}(y_k)\ket{\lambda_2} \\
  & = \sum_{\mu \subseteq [\min(N, N'), M]} s_{\mu/\lambda_1}(\bm{x}) s_{\mu/\lambda_2}(\bm{y}) \; ,
\end{split}
\end{equation}
where \(N' + \ell(\lambda_2) + n_0 = N' + \ell(\lambda_2) + n'_0\) and
we have also used that the skew Schur polynomial for any Young diagram
\(\mu\) that does not contain \(\lambda_1\) and/or \(\lambda_2\)
vanishes.

In the limit \(M, N, N' \to \infty\), we can use the elementary
properties of skew Schur polynomials~\cite{Macdonald:1998}
\begin{equation}
  \sum_{\mu} s_{\mu/\lambda_1}(\bm{x}) s_{\mu/\lambda_2}(\bm{y}) = \prod_{i,j}\frac{1}{1 - x_i y_j}
 \sum_\nu s_{\lambda_1/\nu}(\bm{x}) s_{\lambda_2/\nu}(\bm{y})\; ,
\end{equation}
and the Cauchy's identity, we have that  
\begin{equation}
\begin{split}
  \lim_{M, N, N'\to \infty} A_{\lambda_1 \lambda_2}(N', N, M|\bm{x}, \bm{y})
  & = \sum_\mu s_{\mu}(\bm{x}) s_{\mu}(\bm{y})
 \sum_\nu s_{\lambda_1/\nu}(\bm{x}) s_{\lambda_2/\nu}(\bm{y})\\ 
  & = \sum_\mu s_{\mu}(\bm{x}) s_{\mu}(\bm{y})
 \sum_\nu s_{\lambda_1/\nu}(\bm{x}) s_{\lambda_2/\nu}(\bm{y})\; .
\end{split}
\end{equation}
Therefore, we write the finite case as
\begin{equation}
  A_{\lambda_1 \lambda_2}(N', N, M|\bm{x}, \bm{y})
  = \mathcal{I}(\min(N',N),M, \bm{x}, \bm{y})
  \sum_\nu s_{\lambda_1/\nu}(\bm{x}) s_{\lambda_2/\nu}(\bm{y})\; .
\end{equation}
We observe that this correlation function can be expressed as the
product of the off-shell norm~(\ref{eq:scalar}) and a finite sum over
skew Schur functions.


\subsubsection{Schur polynomials expansion of some correlation functions}
Indeed, it is worth noting that other quantities, which are not tau
functions, may have interesting Schur polynomial expansions. Consider
the state calculated in~\cite{Bogoliubov2005}
\begin{equation}
  \ket{\mathcal{Y}} =
 \sum_{\vec{n}} \prod_{j=0}^M \ket{n_j} = \sum_\mu \ket{\mu} \qquad \sum_j n_j = N\; .
\end{equation}
Then
\begin{equation}
 \bra{\mathcal{Y}} \prod_{j=1}^N B(\bm{x}) \ket{\nu} = 
  \sum_{\lambda\subseteq [N,M]} s_{\lambda/\nu}(\bm{x})\; . 
\end{equation}
While it may not be a tau function, it has an interesting expansion as
a sum of skew Schur polynomials.


\subsection{General tau functions}
Based on the discussion we have had so far, and on the general mapping
between the phase model and free fermions, one can grasp the general
form of tau functions in the context of this integrable chain.

Let us consider the vertex operator construction~\cite{Okounkov2001},
as also discussed in~\cite{Alexandrov:2012tr, Wheeler:2010vmq}. We
consider the vacuum state \(\ket{\bm{0}}\), often referred to as a
``Fermi sea'', defined by the conditions \(\psi_m\ket{\bm{0}} =
\psi_n^\star\ket{\bm{0}} =0 \) for \(m<0\) and \(n \geq 0\), where
\(\psi_n\) are components of a holomorphic free fermionic field. In
this formalism, the partition states are given by
\begin{equation}
  \ket{\mu} = \textrm{sign}(\sigma) \prod_{j=1}^d \psi_{a_j} \psi^\star_{-b_j}\ket{\bm{0}}\; ,
\end{equation}
where \(\textrm{sign}(\sigma) = \pm 1\) are defined in a such a way
that the Shur coefficients of the vertex operators (defined below)
have positive coefficients.

The pairs \(\{(a_j|b_j)\}_{j=1}^d\) define the Frobenius notation of
the partition \(\mu = (\mu_1, \mu_2, \dots, \mu_\ell)\). In this
notation, \(a_j\) is given by \(\mu_j - j\) and \(b_j\) is given by
\(\mu'_j - j\), where \(d\) represents the number of boxes in the
diagonal of the Young diagram, and \(\mu'\) is its conjugate, or
transpose, diagram. From these definitions, we have the equivalence
\begin{equation}
  \prod_{k\geq 1} (\phi_k^\dagger)^{n_k} \mapsto \textrm{sign}(\sigma) \prod_{j=1}^d
  \psi_{a_j} \psi^\star_{-b_j} \qquad |\mu| = \sum_k k n_k = \sum_j(a_j + b_j) + d\; .
\end{equation}

Finally, the \(\mathfrak{gl}(\infty)\) algebra has generators given by the
bilinears \(X = \sum_{j, j \in \mathbb{Z}} x_{ij} \bm{\colon} \psi_i \psi_j^\star\bm{\colon}
+ c\), where \(c\in \mathbb{C}\), \(x_{ij} =
0\) for large \(|j -i|\), say \(\geq M\), and the colons denote the normal ordering
\begin{equation}
  \bm{\colon} \psi_i \psi_j^\ast \bm{\colon} =  \psi_i \psi_j^\ast
  - \bra{0} \psi_i \psi_j^\ast \ket{0} \; .
\end{equation}
The group \(GL(\infty)\) is defined through the exponential map
\(\bm{\exp}: \mathfrak{gl}(\infty) \to GL(\infty)\) as usual.

Note that these elements have only finitely many non-zero entries: the
diagonal terms represent number operators \(\mathcal{N}\), the upper
triangular terms represent annihilation operators \(\phi\), and the
lower triangular terms represent creation operators
\(\phi^\dagger\). The central charge corresponds to the vacuum
projection \(\pi = \ket{\bm{0}}\bra{\bm{0}}\).  Therefore,
\begin{equation}
   GL(\infty) \ni G \mapsto
   \mathcal{G} = \exp \left[\sum_{i=1}^M \left( \sum_{a=1}^3
   c_{i, a} T_i^{a}  + c \pi_i \right)\right] \; \in  \mathrm{Aut}(\mathcal{F})\; ,
\end{equation}
where \(T^{a}_i \in \{ \mathcal{N}_i, \phi_i, \phi_i^\dagger \}_{i=1}^M\). 

The operators \(\mathbb{B}(x)\) and \(\mathbb{C}(x)\), 
for a large enough chain \(M\to \infty\),
are related to the vertex operators
\(\Gamma_-(x)\) and \(\Gamma_+(x)\), respectively, as
\begin{equation}
    \mathbb{B}(x) \mapsto \Gamma_-(x)  = \exp \left( \sum_{n\geq 1} \frac{1}{n}x^n J_{-n}\right) \qquad 
    \mathbb{C}(x) \mapsto \Gamma_+(x)  = \exp \left( \sum_{n\geq 1} \frac{1}{n}x^n J_{n}\right) \; ,
\end{equation}
where \(J_n \) is written in terms of free fermions as \(J_n =
\sum_{j\in \mathbb{Z}} \bm{\colon} \psi_j \psi_{j+n}^\ast
\bm{\colon}\). This set of operators generates a Heisenberg subalgebra
\(\widehat{\mathfrak{gl}}(1) \subset \mathfrak{gl}(\infty)\)
\begin{equation}
  [J_m, J_n] = m \delta_{n+m,0}\; .
\end{equation}

Putting all these facts together, we have that the tau functions of the Toda 
hierarchies, given by
\begin{equation}
  \tau_s(\bm{x}, \bm{y}) = \bra{s} \prod_{i} \Gamma_+(x_i) G \prod_j \Gamma_-(x_j) \ket{s}\; ,
\end{equation}
are mapped into objects of the form 
\begin{equation}
\label{eq:tau-corr}
\begin{split}
  \tau(\bm{x}, \bm{y}) & = \bra{\Psi(\bm{x})} \mathcal{G} \ket{\Psi(\bm{y})} \\
  & = \bra{\bm{0}} \prod_{i} \mathbb{C}(x_i)
  \mathcal{G} \prod_j \mathbb{B}(y_j) \ket{\bm{0}}\; ,
\end{split}
\end{equation}
where we necessarily have \(s=0\) in the phase model


\subsubsection{Hypergeometric tau functions}
From these expressions, we can conclude that if we consider a diagonal
group element \(\mathcal{G} = \exp \left( \sum_{i\geq 0} c_i
\mathcal{N}_i\right)\), we have that~(\ref{eq:tau-corr}) becomes
\begin{equation}
\begin{split}
  \tau(\bm{x}, \bm{y}) 
  & = \bra{\bm{0}} \prod_{i} \mathbb{C}(x_i)
  e^{\sum_i c_i \mathcal{N}_i} \prod_j \mathbb{B}(y_j) \ket{\bm{0}}\\
  & = \sum_{\mu, \nu \subseteq [N,M]} c_{\mu\nu} s_\mu(\bm{x}) s_\nu(\bm{y})
\end{split}
\end{equation}
where
\begin{equation}
  c_{\mu \nu} = \bra{\mu} e^{\sum_i c_i \mathcal{N}_i}\ket{\nu} \; .
\end{equation}
From the representation of the phase model, we have that for the partition
\(\mu = (1^{n_1} 2^{n_2} \dots M^{n_M})\), we have
\begin{equation}
  \mathcal{N}_i \ket{\mu} = n_i \ket{\mu} \qquad n_0 = N - \ell(\mu)\; .
\end{equation}

Consequently, we have
\begin{equation}
  c_{\mu \nu} \equiv \delta_{\mu\nu} c_{\mu}  =  \delta_{\mu\nu} \prod_i e^{c_i n_i}\; , 
\end{equation}
and we conclude that 
\begin{equation}
  \tau(\bm{x}, \bm{y}) 
  = \sum_{\mu, \nu \subseteq [N,M]} c_{\mu} s_\mu(\bm{x}) s_\mu(\bm{y})\; .
\end{equation}
We observe that this tau function has diagonal coordinates
\(c_\lambda\). These tau functions belong to the hypergeometric type
considered in~\cite{Orlov:2000, orlov:2001, Orlov:2001b, Orlov:2005}.

\section{Tau functions in the Q-bosons model}

We now shift our focus to the case of q-bosons. The analysis parallels
what we did before, but the specific details are markedly
different. We begin by examining the norm of two off-shell Bethe
states.

It has been shown~\cite{Tsilevich:2006} (see
also~\cite{Sulkowski:2008mx, Wheeler:2010vmq}) that the Bethe states
\(\ket{\Psi(\bm{x})}\) have coordinate expansions
\begin{equation}
  \ket{\Psi(\bm{x})} = \sum_{\mu \subseteq [N,M]} P_\mu(\bm{x}, Q) \ket{\mu}_Q  \qquad 
  \bra{\Psi(\bm{x})} = \sum_{\mu \subseteq [N,M]} P_\mu(\bm{x}, Q) \prescript{}{Q}{\bra{\mu}}\; ,
\end{equation}
where \(P_\lambda\) denote Hall-Littlewood polynomials. In this form,
the scalar product of two off-shell Bethe states in the q-boson model
can be easily calculated to be
\begin{equation}
\label{eq:scalar-hl}
\mathcal{I}_Q(N,M | \bm{x}, \bm{y}) = \bra{\bm{0}} \prod_{j=1}^N \mathbb{C}(x_j, Q)
\prod_{k=1}^N \mathbb{B}(y_k, Q) \ket{\bm{0}}
= \sum_{\lambda \subseteq [N,M]} b_\lambda(Q) P_{\lambda}(\bm{x}, Q) P_{\lambda}(\bm{y}, Q)
\end{equation}
where we use that \(\bracket{\lambda}{\mu}_Q = b_\lambda(Q)
\delta_{\lambda, \mu}\) and the completeness relation
\begin{equation}
  \sum_{\lambda} \frac{1}{b_\lambda(Q)} \ket{\lambda}_Q {}_Q\bra{\lambda}  = \mathbb{1}\; .
\end{equation}

It turns out that this expansion is the Cauchy identity for
Hall-Littlewood polynomials~\cite{Macdonald:1998}
\begin{equation}
\label{eq:cauchy_hl}
\sum_{\lambda} b_\lambda(Q) P_{\lambda}(\bm{x}, Q) P_{\lambda}(\bm{y}, Q)
= \prod_{j, k=1}^\infty \frac{1-Q x_j y_k}{1 - x_j y_k}\; .
\end{equation}

Our goal is to explore some properties of Hall-Littlewood polynomials
to gain insight into aspects of this expansion.


\subsection{Scalar product: determinant formula}
We aim to refine the expressions above. First, we consider a
determinant expression for the scalar product~(\ref{eq:scalar-hl}). We
proceed with the scenario where \(M\) and \(N\) are very large but
finite. In this case, we express this scalar product as
\begin{equation}
\label{eq:q-inner}
  \mathcal{I}_Q(N,M | \bm{x}, \bm{y})  
= \prod_{j_1, j_2} (1-Q x_{j_1} y_{j_2}) \prod_{k_2, k_2}(1 - x_{k_1} y_{k_2})^{-1}\; .
\end{equation}

Using the Cauchy's identity for Schur polynomials, that is 
\begin{equation}
  \sum_\lambda s_\lambda(\bm{x}) s_\lambda(\bm{y}) = \prod_{i,j} \frac{1}{1 - x_i y_j}\; ,
\end{equation}
and from the results derived in the phase model, we find that
\begin{equation}
  \prod_{i,j}\frac{1}{1 - x_i y_j}  \equiv \mathcal{I}(N,M|\bm{x}, \bm{y}) = 
  \frac{\det_{j,k}H(\bm{x},\bm{y})}{\Delta(\bm{x}) \Delta(\bm{y})}\; ,
\end{equation}
where \(H\) is the matrix~(\ref{eq:h-matrix}). Consequently, the scalar
product~(\ref{eq:q-inner}) becomes
\begin{equation}
  \mathcal{I}_Q(N,M | \bm{x}, \bm{y})  
= \frac{\mathcal{I}(N,M|\bm{x}, \bm{y})}{\mathcal{I}(N,M|\bm{x}, Q\bm{y})}\; , 
\end{equation}
and we see that it is the quotient of scalar products of the phase
model. Hence, it is the quotient of two Toda tau functions.
Observe that in the case \(Q = 0\), we have \(\mathcal{I}(N, M|
\bm{x}, \bm{0}) = 1\), then \(\mathcal{I}_0(N,M | \bm{x}, \bm{y})
= \mathcal{I}(N,M | \bm{x}, \bm{y})\), as expected.

Additionally, we write
\begin{equation}
  \mathcal{I}_Q(N,M | \bm{x}, \bm{y})  
  = \frac{\Delta(Q\bm{y})}{\Delta(\bm{y})}
  \frac{\det H(\bm{x}, \bm{y})}{\det H(\bm{x},Q \bm{y})} \; .
\end{equation}
Using \(\Delta(Q\bm{y})/ \Delta(\bm{y}) = \prod_{j=1}^{N-1} Q^j\), we have 
\begin{equation}
  \mathcal{I}_Q(N,M | \bm{x}, \bm{y}) =  Q^{N(N-1)/2}\;
  \frac{\det H(\bm{x}, \bm{y})}{\det H(\bm{x}, Q \bm{y})} \; .
\end{equation}
Let us denote \(H(\bm{x}, Q\bm{y}) = H_{Q}\), therefore
\begin{equation}
  \mathcal{I}_Q(N,M | \bm{x}, \bm{y}) = Q^{N(N-1)/2} \det H_{Q}^{-1} \det H
= Q^{N(N-1)/2} \det \mathcal{H}(\bm{x}, \bm{y}) \; ,
\end{equation}
where \(\mathcal{H} = H_{Q}^{-1} H\) is a matrix with components 
\(\mathcal{H}_{i,j} \equiv \mathcal{H}(x_i, y_j)\) where
\begin{equation}
  \mathcal{H}(z, w) = \sum_{k} \frac{(1 - z y_k)}{(1 - Q z y_k)} 
\frac{(1 - Q x_k w)}{(1 - x_k w)}\; .
\end{equation}
Note that from this expression, we cannot decompose this function as
in~(\ref{eq:h-exp}) since the coefficients in this expansion depend on
\(\bm{x}\) and \(\bm{y}\).


\subsection{Big Schur functions expansion}
Now, let's revisit the result originally derived in~\cite{Foda:2008hn}
which demonstrates that the scalar product~(\ref{eq:scalar-hl}) is a
tau function of the KP hierarchy. According to~\cite[Chapter 3,
  Section 4, Equation (4.7)]{Macdonald:1998}, we can expand the Cauchy
identity~(\ref{eq:cauchy_hl}) as
\begin{equation}
 \prod_{j, k=1}^\infty \frac{1-Q x_j y_k}{1 - x_j y_k} = 
\sum_{\lambda} \mathcal{S}_{\lambda}(\bm{x}, Q) s_{\lambda}(\bm{y}) =
\sum_{\lambda} \mathcal{S}_{\lambda}(\bm{y}, Q) s_{\lambda}(\bm{x}) \; .
\end{equation}
Here, the polynomials \(\mathcal{S}_\lambda\), which we will refer to as
the big-Schur functions, are defined by a Jacobi-Trudi formula:
\begin{equation}
  \mathcal{S}_{\lambda} (\bm{y}, Q) = \det(q_{\lambda_i -i + j}(\bm{y}, Q))\; , 
\end{equation}
where the coefficients \(q_m\) are obtained from the expression:
\begin{equation}
 \sum_{m} q_m(\bm{y}, Q) z^m =
 \prod_{j} \frac{1-Q y_j z}{1 - y_j z}\; ,
\end{equation}
and \(z\) is a formal variable. It has been
argued~\cite{Foda:2008hn} in that if we interpret the big-Schur
functions as Plücker coordinates, the expression
\begin{equation}
\label{eq:big-expansion}
  \mathcal{I}_Q(N,M | \bm{x}, \bm{y})
  = \sum_{\lambda \subseteq [N,M]} \mathcal{S}_{\lambda}(\bm{x}, Q) s_{\lambda}(\bm{y})
  = \sum_{\lambda \subseteq [N,M]} \mathcal{S}_{\lambda}(\bm{y}, Q) s_{\lambda}(\bm{x})\; ,
\end{equation}
is a restricted KP tau function with respect to both set of coordinates,
that is \(\bm{x}\) and \(\bm{y}\).

Based on these findings, we conclude that the inner product can be
expressed as a quotient of two Toda tau functions or as a KP tau
function with coefficients given by the big Schur polynomials.
Nevertheless, further investigation of these results is necessary.


\subsection{Kostka-Foulkes expansion}
In a Schur polynomial basis, we write
\begin{equation}
P_\lambda(\bm{x}, Q) = \sum_{\mu} K^{-1}_{\lambda \mu}(Q) s_\lambda(\bm{x})\; , 
\end{equation}
where \(K^{-1}_{\mu\nu}(Q) \in \mathbb{Z}[Q]\) are inverse
Kostka-Foulkes polynomials~\cite{Macdonald:1998, Wheeler:2018}.

If we now insert this expansion in the inner product, we have 
\begin{equation}
    \sum_\lambda b_\lambda(Q) P_\lambda(\bm{x}, Q) P_\lambda(\bm{y}, Q) = 
    \sum_{\mu,\nu} \left(\sum_\lambda K_{\mu \lambda}^{-1 \; T} b_\lambda(Q)  K_{\lambda \nu}^{-1} \right)
    s_{\mu}(\bm{x}) s_{\nu}(\bm{y}) 
\end{equation}
and we have the coefficients 
\begin{equation}
\label{eq:c-expan}
 \tilde{c}_{\mu\nu}(Q) = \sum_\lambda K_{\mu \lambda}^{-1 \; T} b_\lambda(Q)  K_{\lambda \nu}^{-1} \; ,
\end{equation}
for the double expansion of the inner product in terms of Schur
polynomials. Based on our discussion so far, we know that these
coefficients are not Plücker coordinates of the Toda hierarchy.

But we can say something interesting about these coefficients. Let us
now expand the big-Schur functions as
\begin{equation}
\label{eq:kf-exp}
  \mathcal{S}_\mu(\bm{x}, Q) = \sum_{\lambda} \mathcal{C}^{\mu}_\lambda(Q) s_\lambda(\bm{x})\; ,
\end{equation}
we can fix the coefficients \(\mathcal{C}_\lambda^\mu(Q)\) in terms of
Kostka-Foulkes polynomials using the orthogonality relations of these
polynomials~\cite{Macdonald:1998}. In particular, there is an inner
product in the ring of symmetric functions such that
\begin{equation}
  b_\mu \langle P_\mu(\bm{x}, Q), P_\nu(\bm{x}, Q)\rangle = \langle
  \mathcal{S}_\mu(\bm{x}, Q), s_\nu(\bm{x})\rangle = \delta_{\mu\nu}\; .
\end{equation}
Therefore,
\begin{equation}
  \begin{split}
\delta_{\mu\nu} & = \sum_\lambda \mathcal{C}_\lambda^\mu \langle s_\lambda(\bm{x}), s_\nu(\bm{x})\rangle
 = \sum_\lambda \sum_{\pi, \sigma} \mathcal{C}_\lambda^\mu K_{\lambda \pi} K_{\nu \sigma}
\langle P_{\pi}(\bm{x}, Q), P_{\sigma}(\bm{x}, Q)\rangle \\ 
& = \sum_\lambda \mathcal{C}_\lambda^\mu \sum_{\sigma}K_{\lambda \sigma} b_{\sigma}^{-1} K_{\nu \sigma}\; ,
  \end{split}
\end{equation}
and we conclude that 
\begin{equation}
\label{eq:indices}
(\mathcal{C}^{-1})^\lambda_\nu = \sum_\sigma K_{\lambda\sigma} b^{-1}_\sigma (K_{\sigma\nu})^T\; ,
\end{equation}
and consistency with~(\ref{eq:c-expan}) implies that
\(\mathcal{C}^\lambda_\nu \equiv \tilde{c}_{\lambda\nu}\).

Hence, we conclude that 
\begin{equation}
S_{\mu}(\bm{t},Q) = \sum_{\nu} \tilde{c}_{\mu\nu}(Q) s_{\mu}(\bm{t})\; ,
\end{equation}
where we have written the big Schur polynomial in terms of the Miwa
coordinates \(\bm{t}\).

\begin{remark}
This expression also reveals something interesting. We can formulate
this problem in terms of partition states defined in the phase
model. Alternatively, we can utilize the conventional vertex operator
construction rather than the q-deformed version proposed by
Jing~\cite{Jing1991, Jing1995}, which is much more challenging to
handle.
\end{remark}


\subsection{Supersymmetric Schur polynomials expansion}
Based on this result, we argue that although Equation
\eqref{eq:scalar-hl} is not a Toda tau function with respect to the
coordinates \( \{\bm{t}; \bm{t}'\} \), we can define a new set of
coordinates \( \{\bm{t};\bm{T}\} \) such that the scalar product
becomes a trivial Toda tau function.

Let us first decompose
\begin{equation}
 \sum_{m} q_m(\bm{y}, Q) z^m =
 \prod_j (1 + y_j (-Q z)) \prod_k (1 - y_k z)^{-1} = 
 \sum_{j, k} e_j(\bm{y}) h_k(\bm{y}) (- Q)^j z^{j+k}\; .
\end{equation}
Reorganizing this sum, we conclude that
\begin{equation}
  q_m(\bm{y}, Q)  = \sum_{j=0}^m e_j(\bm{y}) h_{m-j}(\bm{y}) (- Q)^j =
 \sum_{j=0}^m e_j(-Q\bm{y}) h_{m-j}(\bm{y}) \; ,
\end{equation}
where in the last equality we have used the homogeneity of the
elementary symmetric polynomials, and \( Q\bm{y} \equiv (Qy_1, Qy_2,
\dots)\).

It is easy to see that
\begin{subequations}
\begin{equation}
  \begin{split}
    \prod_{j=1}^N (1 - (Q y_j)z) & = \prod_{j=1}^N e^{ \ln  (1 - (Q y_j)z)} = 
    \prod_{j=1}^N \exp \left( - \sum_{n>0} \frac{Q^n y^n_j}{n} z^n \right) \\ 
    & = \exp \left( - \sum_{n>0} \sum_{j=1}^N \frac{Q^n y^n_j}{n} z^n \right)  =
    \exp \left( - \sum_{n>0} t^{(Q)}_n z^n \right)  
  \end{split}
\end{equation}
where \(t^{(Q)}_n = \frac{Q^n}{n} \sum_j y_j^n = Q^n t'_{-n}\). Additionally
\begin{equation}
    \prod_{j=1}^N \frac{1}{(1 - y_jz)} = \prod_{j=1}^N e^{- \ln  (1 - y_jz)} = 
    \prod_{j=1}^N \exp \left( \sum_{n>0} \frac{y_j}{n} z^n \right)=
    \exp \left(\sum_{n>0} t'_{-n} z^n \right) \; . 
\end{equation}
\end{subequations}

Consequently
\begin{equation}
 \sum_{m} q_m(\bm{y}, Q) z^m =
 \prod_j \frac{1 - Q y_j z}{1 - y_k z}
 = \exp \left( \sum_{n\geq 1} (t'_{-n} - t_n^{(Q)})z^n \right)
 \equiv \exp \left( \sum_{n\geq 1} T_n z^n \right)\; ,
\end{equation}
with \(T_n = (1 - Q^n) t'_{-n}\). Then, we conclude that \(q_m(\bm{y}, Q)\)
are homogeneous polynomials with respect the Miwa coordinates
\(\bm{T} = (T_1, T_2, \dots)\). Then
\begin{equation}
  \mathcal{S}_\lambda(\bm{t}', Q) = \det \left(h_{\lambda_i - i +j}(\bm{T})\right) = s_\lambda(\bm{T})\; .
\end{equation}
All in all, we conclude that the big Schur functions
\(S_\lambda(\bm{t}', Q)\) are ordinary Schur functions with respect to
the coordinates \(\bm{T}\).

Moreover, a more refined approach is also possible. Supersymmetric (or
Hook) Schur functions~\cite{Berele:1983}, as discussed in works such
as~\cite{Macdonald:1998, Moens:2003}, denoted by
\(s_\lambda(\bm{\alpha}/\bm{\beta})\), are defined as ordinary Schur
functions evaluated at Miwa coordinates of the form
\begin{equation}
  T_n = \frac{1}{n} \left( \sum_{i=1}^{\dim(\alpha)} \alpha_i^n
       - \sum_{i=1}^{\dim(\beta)} (-\beta_i)^n\right)\; .
\end{equation}
Comparing this expression with the results above, we can see that the
big Schur functions \(S_\lambda(\bm{t}', Q)\) correspond to the
supersymmetric Schur functions for \(\bm{\alpha} = \bm{y}\) and
\(\bm{\beta} = - Q \bm{y}\), which is
\begin{equation}
  S_\lambda(\bm{t}', Q) = s_\lambda[\bm{y}/(- Q\bm{y})]\; .
\end{equation}

Putting these facts together, we immediately conclude that
\begin{equation}
\lim_{M,N\to \infty}\mathcal{I}_Q(N,M|\bm{x}, \bm{y}) = 
  \sum_\lambda s_\lambda(\bm{T}) s_\lambda (\bm{t})
\end{equation}
is a Toda hierarchy tau function with respect to
\(\{\bm{t};\bm{T}\}\). As a result, we obviously have that
\begin{equation}
  \mathcal{I}_Q(N,M| \bm{x}, \bm{y}) \equiv \mathcal{I}_Q(N,M| \bm{t}, \bm{T})
  = \sum_{\lambda \subseteq [N,M]} s_\lambda(\bm{T}) s_\lambda(\bm{t})
\end{equation}
is also a restricted tau function of the Toda hierarchy with respect
to \(\bm{t}\) and \(\bm{T}\).

Let us also express the supersymmetric Schur polynomials in terms of
ordinary Schur functions, as shown in~\cite[Sec. I.5, exerc. 23]{Macdonald:1998}:
\begin{equation}
  s_{\lambda}(\bm{x}/\bm{y}) = \sum_\mu s_{(\lambda/\mu)'}(\bm{y}) s_\mu(\bm{x})\; ,
\end{equation}
where the prime denotes the conjugate diagram. Compare this expansion
with~(\ref{eq:kf-exp}). Then
\begin{equation}
  \label{eq:tau-hl}
  \sum_{\lambda } s_\lambda(\bm{T}) s_\lambda(\bm{t}) =
  \sum_{\lambda,\mu } s_{(\lambda/\mu)'}(- Q\bm{y}) s_\lambda(\bm{t}) s_\mu(\bm{t}')\; .
\end{equation}

From this expression we conclude (and speculate) the following. 

\begin{remark}
Since the skew Schur polynomials have determinant expressions, we can
deduce that the \(\bm{y}\)-dependent coefficients \(c_{\mu \lambda} =
s_{(\lambda/\mu)'}(-Q\bm{y})\) have Jacobi-Trudi expressions.  It is
tempting to regard these objects as \(\bm{y}\)-dependent Plücker
coordinates.  In this sense, we would have a curve in the infinite
Grassmannian instead of a point. Evidently, it is not a tau functions
on any known integrable hierarchy, but it might suggest some new
generalizations that are worth investigating.
\end{remark}

\begin{remark}
It is worth noting that one of the simplest nontrivial solutions of
the KP hierarchies are the Schur functions
themselves~\emph{\cite{Zabrodin2018}}. As we have just demonstrated, the big
Schur polynomials can be expressed as supersymmetric Schur
polynomials, which are essentially ordinary Schur polynomials in a
specific choice of Miwa coordinates. Therefore, we conclude that the
big Schur polynomials are also KP tau functions. This direct
conclusion serves as an alternative proof of this fact, originally derived
in~\emph{\cite{Necoechea:2019wbg}} using the KP bilinear identity.
\end{remark}

Combining these observations, we conclude that the left-hand side of
equation \(\ref{eq:tau-hl}\) is a tau function with respect to
the coordinates \(\{\bm{t}; \bm{T}\}\). Furthermore, by employing the
vertex operator formalism, we can express the Schur polynomials as
\begin{equation}
  s_{\lambda}(\bm{t}') = \bra{\lambda} e^{J_-(\bm{t}')}\ket{\bm{0}}\qquad
  s_{\mu}(\bm{t}) = \bra{\bm{0}} e^{J_+(\bm{t})}\ket{\mu}\; . 
\end{equation}
Hence
\begin{equation}
  \sum_{\lambda,\mu } c_{\mu\lambda}(Q,\bm{y}) s_\mu(\bm{t}') s_\lambda(\bm{t}) = 
  \sum_{\lambda,\mu } \bra{\bm{0}} e^{J_+(\bm{t})}\ket{\mu} c_{\mu\lambda}(Q,\bm{y})
  \bra{\lambda} e^{J_-(\bm{t}')}\ket{\bm{0}} \; .
\end{equation}
It is important to reiterate that this expression does not constitute
a tau function with respect to \( \{\bm{t}, \bm{t}'\} \). Despite this
crucial distinction, we proceed under the assumption that the
coefficients can be expressed as
\begin{equation}
  c_{\mu\lambda}(Q, \bm{y}) = \bra{\mu} G_Q(\bm{y}) \ket{\lambda}\; ,
\end{equation}
where \(G_Q(\bm{y}) \in GL(\infty)\). Hence
\begin{equation}
  \lim_{N,M\to\infty} \mathcal{I}_Q(N, M| \bm{t}, \bm{t}'(\bm{y})) = 
  \bra{\bm{0}} e^{J_+(\bm{t})} G_Q(\bm{y}) e^{J_-(\bm{t}')}\ket{\bm{0}}\; ,
\end{equation}
where \(t'_{-n} = \frac{1}{n}\sum_i y_i^n\). 

To derive the expression for this group element, recall that \(T_m =
t'_{-m} - Q^m t'_{-m}\). With this in mind, we can express the scalar
product as follows:
\begin{subequations}
\begin{equation}
  \bra{\bm{0}} e^{J_+(\bm{t})} e^{J_-(\bm{T})} \ket{\bm{0}} = 
\bra{\bm{0}} e^{J_+(\bm{t})} e^{J_-(\bm{t}')  - J_-(\bm{t}^{(Q)})} \ket{\bm{0}} \; .
\end{equation}
From the Heisenberg algebra, we deduce that \([J_-(\bm{t}),
  J_-(\bm{t}')] = 0\), therefore
\begin{equation}
  \bra{\bm{0}} e^{J_+(\bm{t})} e^{J_-(\bm{T})} \ket{\bm{0}} = 
\bra{\bm{0}} e^{J_+(\bm{t})} e^{-J_-(\bm{t}^{(Q)})} e^{J_-(\bm{t}')}  \ket{\bm{0}} \; .
\end{equation}
\end{subequations}
We finally conclude that 
\begin{equation}
  G_{Q}(\bm{y}) = \exp \left(-J_-(\bm{t}^{(Q)})\right) \qquad
  t^{(Q)}_n = \frac{Q^n}{n}\sum_j y_j^n\; . 
\end{equation}

Generally, the expectation values involving coordinate-dependent group
elements \(G_{Q}(\bm{y})\) do not form Toda hierarchy tau
functions. However, in the example above, the coordinates combine in
such a way that they generate a tau function with respect to the
coordinates \(\bm{T}\) and \(\bm{t}\).
 

\section{Discussion, Conclusions and Perspectives}

In this work, we have explored the connections between a quantum
integrable system, the q-boson model, and solutions of a classical
integrable system, the Toda hierarchy. Our investigation has extended
some early findings in this field and has also unveiled new research
avenues, which we aim to explore in future studies. Let us now discuss
some of these promising directions.

In Section 3, we explored various aspects of the phase model,
presenting the scalar product of two off-shell Bethe states as an
elementary example of a Toda system tau function. This section
overlaps with existing literature, particularly a paper by Wheeler and
Foda~\cite{Wheeler:2010vmq, Foda:2008hn}. In their work, they
demonstrate that the inner products of two Bethe states are KP
hierarchy tau functions. Our current work builds on this by arguing
that these scalar products are also Toda hierarchy tau functions, as
they can be expanded as a Cauchy identity, with the coefficients being
Schur functions. When one set of coefficients is fixed in this
expression, Wheeler's results are recovered.

We also extend this analysis to other correlation functions,
determining when these functions satisfy integrable hierarchy
equations. Using expressions derived by Bogoliubov and
collaborators~\cite{Bogoliubov:1997soj}, we reformulated them to check
if the correlation functions are KP/Toda tau functions. Although these
objects cannot generally be written as Toda tau functions, with some
adjustments, they can be expressed as KP tau functions.  We have
elucidated how various correlation functions align with this
framework, revealing a remarkably rich structure.

The main results are discussed in Section 3.3, where we examine the
general form of tau functions associated with this phase model.
Notably, we explored a mapping between the phase model
data and the vertex operator representation of free bosons. An
intriguing avenue for future research is to delve into the properties
of the hypergeometric tau functions uncovered in Section 3.3.1.
Exploring these functions is expected to provide significant
understanding, and comparing them with existing literature will
further enrich our comprehension of their characteristics and
implications within the context of integrable systems. This problem is
currently under investigation. 

Additionally, we revealed an intriguing alternative portrayal of these
scalar products using a matrix integral framework, which corresponds
to an ensemble of Coulomb particles. This discovery opens up promising
avenues for further inquiry. Specifically, it would be intriguing to
further explore this subject and investigate whether the matrix model
description provides valuable insights into the phase model and its
Bethe roots. Such an investigation promises to illuminate the
underlying dynamics of the phase model and its relationship with
classical integrable systems.

In Section 4, we addressed the same problem within the context of the
q-boson model, attempting to replicate the analysis done in Section
3. There are few known results in the literature regarding the
relationship between these objects and KP/Toda tau functions, so we
used established formulas to investigate if the scalar products are
also tau functions.

Initially, we derived a determinant formula for the q-boson scalar
products and discussed their expression in terms of the phase model
data. This result establishes a connection between the q-boson and
phase model quantities. Furthermore, we explored different expansions
for these scalar products. Notably, we observed that they can be
expanded in terms of Big Schur and supersymmetric tau
functions. Consequently, we introduced a new set of coordinates,
demonstrating that scalar products can be precisely expressed as Toda
tau functions within this transformed coordinate system.

We also write these formulas in terms of simpler orthogonal
polynomials. Firstly, we show that these scalar products cannot be
naively written as Toda tau functions. However, using some known
expansions, we expand the inner products in different forms. In one
of these expansions, we define a new set of coordinates, making the
scalar product of q-bosons Toda tau functions with respect to these
new coordinates. This is the main result of this section.

One of our most pressing challenges lies in elucidating the intricate
connection between our findings and the Ablowitz-Ladik
hierarchy. Since this hierarchy arises as a reduction of the Toda
hierarchy, it becomes imperative to understand how we can capture the
structure of the AL hierarchy using the results we have derived.
Considering that the q-boson model effectively quantizes the
Ablowitz-Ladik equation, it follows that we should detect some
resemblance of this classical problem within the q-boson system.  This
understanding holds the promise of shedding significant light on the
interplay between classical and quantum integrable systems.

We hope to address some of these challenges in future publications.

\subsubsection*{Acknowledgments}
This work is supported by Fapesp through grant \textbf{2022/06599-0}.

\printbibliography

\end{document}

%% file: definitions.tex
\usepackage{parskip}
\usepackage{tikz} 
\usepackage[backend=biber, style=alphabetic]{biblatex}
\usepackage{listings}
\usepackage{ytableau}

\usetikzlibrary{intersections,decorations.text} 
\usetikzlibrary{matrix, arrows.meta} 

\hypersetup{ 
	pdftitle={q-Bosons and integrable hierarchies},
	pdfsubject={Mathematical-Physics}, 
	pdfauthor={author},
	pdfkeywords={integrability, bethe, schur, cft},
	colorlinks=true, 
    linkcolor=myPurple, 
    citecolor=myPurple, 
    urlcolor =myPurple, 
    linktocpage=true 
}

\definecolor{myPurple}{RGB}{90, 74, 120}
\definecolor{myBlue}{RGB}{15, 75, 110}
\definecolor{myRed}{RGB}{191,97,106}
\definecolor{myDarkGray}{RGB}{216, 222, 233}
\definecolor{myLightGray}{RGB}{236, 239, 244}

\definecolor{c1}{RGB}{129, 162, 193}
\definecolor{c2}{RGB}{216, 222, 233} 
\definecolor{c3}{RGB}{236, 239, 244} 
\definecolor{c4}{RGB}{59, 66, 82}
\definecolor{c5}{RGB}{76, 86, 106}

\definecolor{shadecolor}{RGB}{236, 239, 244}

\newtheorem{remark}{Remark}

\DeclarePairedDelimiter{\bra}{\langle}{\rvert}
\DeclarePairedDelimiter{\ket}{\lvert}{\rangle}

\DeclarePairedDelimiterX{\bracket}[2]{\langle}{\rangle}{#1\vert#2}
\DeclarePairedDelimiterX{\bbracket}[2]{\langle\!\langle}{\rangle\!\rangle}{#1\vert#2}